\documentclass[12pt]{article}
\usepackage{graphicx}
\usepackage{epstopdf}
\usepackage{amsmath,amssymb,amsfonts,graphicx}

\begin{document}
\thispagestyle{empty}
\renewcommand{\refname}{References}

\title{\bf The Aharonov-Bohm effect in scattering \\ of nonrelativistic electrons by \\ a penetrable magnetic vortex} 

\author{Yurii A. Sitenko}

\date{}

\maketitle

\begin{center}
Bogolyubov Institute for Theoretical Physics, National Academy
of Sciences, \\ 14-b Metrologichna
Str., Kyiv, 03680, Ukraine \\
\end{center}

\begin{abstract}
Quantum-mechanical theory for scattering of nonrelativistic charged particles
with spin by a penetrable magnetic vortex is elaborated. The scattering 
differential cross section is shown to consist of two terms, one describing 
diffraction on the vortex in the forward direction and another one describing 
penetration through the vortex. The Aharonov-Bohm effect is manifested as a 
fringe shift in the diffraction pattern. The penetration effect is analyzed 
for the case of the uniform distribution of the magnetic field strength 
inside the vortex. We find that the penetrability of the magnetic vortex 
does not affect the diffraction pattern, and, hence, the Aharonov-Bohm effect 
is the same for a penetrable vortex as for an impenetrable one. 
\end{abstract}


\bigskip

\begin{center}
\end{center}

\bigskip

\newpage

\section{Introduction}

The theoretical prediction of the Aharonov-Bohm (AB) effect in 1959
\cite{Aha} was one of the most intriguing achievements in quantum
theory. Now this effect has been long recognized for its crucial
role in demonstrating that, in addition to the usual local
(classical) influence of electromagnetic field on charged particles,
there exists the unusual nonlocal (purely quantum) influence of
electromagnetic fluxes confined in the regions which are
inaccessible to charged particles (see, e.g., reviews
\cite{Pes,Ton}). The AB effect is validated in experiments 
on detecting a fringe shift in the
interference pattern due to two coherent electron beams under the
influence of an impenetrable magnetic vortex placed between the
beams. It should be noted that during several decades the concern 
of experimentalists was to ensure the impenetrability of the magnetic 
vortex, since the issue of the overlap between the region of the magnetic 
flux and that accessible to electron beams was the main reason for numerous attempts 
to refute the validation \cite{Pes,Ton}. However, none have acknowledged 
a rather paradoxical circumstance that, {\it post factum}, all these 
efforts turn out to be unnecessary, with all refutations disproved at once: 
experimental data testify that the AB effect is independent of the extent 
of electron penetration into a magnetized sample. For instance, it follows 
from Fig.3 in \cite{Ton} that the fringe shift is quantitatively the same for 
a transparent magnet and an opaque magnet (covered with a gold film which is by 
two orders of magnitude thicker than the magnet itself).

This empirical fact has to be theoretically explained. 
Scattering theory for the AB effect was initiated in seminal paper 
\cite{Aha} (see also \cite{Aha1}), and, later on, it was substantiated 
and further developed in works \cite{Ber,Rui,Au,Jac}. 
So far scattering by a magnetic vortex of zero transverse size was considered, 
while the account for nonzero transverse size of the vortex was properly taken 
in more recent works \cite{Si10,Si11,SiV,Si12,Si13}. A theoretical  
explanation of the above empirical fact will be given in the present paper, basing 
on these previous studies.

Thereafter, we are considering scattering of a nonrelativistic electron (or, somewhat more generally, a 
charged particle with spin) on a magnetic vortex, by which we mean a magnetic field configuration in the form of a long, formally infinite, tube of finite radius. Such a configuration can be created by a long current-carrying solenoid, or, otherwise, it is formed inside a long whisker of a magnetized material. To take account for a partial penetrability of the magnetic vortex, we introduce an infinitely thin potential barrier at its edge: $V(r)=\frac{\hbar^2}{2m}\kappa \delta(r-r_c)$, where $r_c$ is the radius of the vortex and $m$ is the mass of the scattered particle; the case of $\kappa  = \pm \infty$ corresponds to the fully impenetrable vortex. The smallest possible value of the vortex radius can be of order 10$^{-7}$ m \cite{Ton}, while the largest possible values of the charged particle wavelength are of order 10$^{-10}$ m (slowly moving electrons of energies 10-100 eV). Thus, one has for sure
\begin{equation}
kr_c \gg 1,\label{eq1}
\end{equation}
where $k=2\pi /\lambda$ is the value of the particle wave number vector ($\lambda$ is the particle wavelength). The radius of a classical orbit of the charged particle in the uniform magnetic field, $r_B=\hbar kc/|eB|$ ($B$ is the magnetic field strength, $e$ is the particle charge), is assumed to exceed the particle wavelength: $r_B>\lambda$. This restricts the values of the magnetic field strength to be less than its atomic unit, $|B|<10^9$ G, in the case of a scattered particle of the maximally possible wavelength (the bound increases with the decrease of the wavelength). It should be noted that the maximal values of steady magnetic fields which are attainable in laboratory are of order 10$^5$ G, see, e.g., \cite{Per}, and this allows us for sure to impose a stronger restriction: $r_B\gg k^{-1}$. Defining the total flux of the uniform magnetic field, $\Phi=\pi r_c^2B$, we rewrite the latter restriction as
\begin{equation}
|e\Phi|(\pi\hbar c)^{-1}\ll(kr_c)^2.\label{eq2}
\end{equation}
Thus, concerning the present scattering problem, we have one large dimensionless parameter, see (1), while the magnetic field with respect to this parameter can be either weak, $|e\Phi|(\pi\hbar c)^{-1} \leq kr_c$ ($r_B \geq r_c$), or strong, $|e\Phi|(\pi\hbar c)^{-1} > kr_c$ ($r_B < r_c$), but is restricted by condition (2). 

In the next section, basing on the Schr\"{o}dinger equation, we derive the scattering amplitude and differential cross section. The obtained results are discussed and summarized in the concluding section. 

\section{Scattering amplitude and cross section}

Let us start with the Schr\"{o}dinger equation for wave function $\Psi$ of a charged particle with spin 1/2: 
\begin{eqnarray}
{\rm i}\hbar\partial_t\Psi(t,r,\varphi,z)=-\frac{\hbar^2}{2m}\Biggl[
\partial_r^2 + \frac{1}{r} \partial_r+\nonumber \\
+\frac{1}{r^2}\left(\partial_\varphi-\frac{{\rm i}e}{\hbar c}A_\varphi\right)^2+\partial_z^2+\frac{e}{\hbar c}\sigma^3B-\kappa\delta(r-r_c)\Biggr]\Psi(t,r,\varphi,z),\label{eq3}
\end{eqnarray}
where magnetic field $B$ is directed along the $z$-axis in cylindrical $(r,\varphi,z)$ coordinates, and  we assume the field to be nonvanishing at $r<r_c$, static and cylindrically symmetric; the gauge is chosen as ${\bf A}=(0,A_{\varphi}(r),0)$, where $A_\varphi(r)=\int\limits_{0}^{r}{\rm d}rrB$, hence 
\begin{equation}
A_\varphi=\frac{\Phi}{2\pi},\quad r>r_c,\label{eq4}
\end{equation}
$\Phi=2\pi\int\limits_{0}^{r_c}{\rm dr}\,rB$ is the total flux of the magnetic field.
Solutions to (3) correspond to continuous energy $E=\frac{\hbar^2}{2m}({\bf k}^2+k_z^2)$, where ${\bf k}$ and $k_z$ are the wave number vectors in the transverse and longitudinal directions (with respect to the magnetic field). The dependence of a solution on time and longitudinal coordinate is obvious, so one can write
\begin{equation}
\Psi(t,r,\varphi,z)=e^{-{\rm i}Et/\hbar}e^{{\rm i}k_zz}\psi(r,\varphi), \qquad r>r_c\label{eq5}
\end{equation}
and
\begin{equation}
\Psi(t,r,\varphi,z)=e^{-{\rm i}Et/\hbar}e^{{\rm i}k_zz}\tau(r,\varphi), \qquad r<r_c,\label{eq6}
\end{equation}
where wave functions of transverse coordinates obey the matching condition at the edge of the vortex:
\begin{equation}
\psi|_{r=r_c}=\tau|_{r=r_c},\quad \partial_r \psi|_{r=r_c}=\partial_r \tau|_{r=r_c}+\kappa\tau|_{r=r_c}.\label{eq7}
\end{equation}
Decomposing the wave functions into partial waves as 
\begin{equation}
\psi(r,\varphi)=\sum\limits_{n\in \mathbb{Z}}e^{{\rm i}n\varphi}a_n\psi_n(r)\label{eq8}
\end{equation}
and
\begin{equation}
\tau(r,\varphi)=\sum\limits_{n\in \mathbb{Z}}e^{{\rm i}n\varphi}b_n\tau_n(r),\label{eq9}
\end{equation}
where $\mathbb{Z}$ is the set of integer numbers, one can get the following equations for the partial waves in view of (3):
\begin{equation}
[r^{-1}\partial_rr\partial_r-r^{-2}(n-\mu)^2+k^2]\psi_n(r)=0\label{eq10}
\end{equation}
and
\begin{equation}
[r^{-1}\partial_rr\partial_r - r^{-2}(n-\gamma)^2 + \sigma^3r^{-1}(\partial_r\gamma) + k^2]\tau_n(r)=0,\label{eq11}
\end{equation}
where $\mu=e\Phi/(2\pi\hbar c)$ and $\gamma(kr)=eA_\varphi(r)/(\hbar c)$.

Given two linearly independent solutions to (10), denoted by $\psi_n^{(+)}(r)$ and $\psi_n^{(-)}(r)$, a solution to (10), which agrees with condition (7), takes the form
\begin{equation}
\psi_n(r)=\psi_n^{(-)}(r)-\left.\frac{W(\psi_n^{(-)},\tau_n)+\kappa\psi_n^{(-)}\tau_n}{W(\psi_n^{(+)},\tau_n)+\kappa\psi_n^{(+)}\tau_n}\right|_{r=r_c}\psi_n^{(+)}(r),\label{eq12}
\end{equation}
where
$$
W(\tau^{(1)},\tau^{(2)})\equiv\tau^{(1)}(r)\partial_r\tau^{(2)}(r)-\tau^{(2)}(r)\partial_r\tau^{(1)}(r)
$$
is the Wronskian of functions $\tau^{(1)}(r)$ and $\tau^{(2)}(r)$. Note also a relation between coefficients $a_n$ and $b_n$:
\begin{equation}
b_n=\Biggl.a_n\frac{W(\psi_n^{(+)},\,\psi_n^{(-)})}{W(\psi_n^{(+)},\,\tau_n)+\kappa\psi_n^{(+)}\tau_n}\Biggr|_{r=r_c}.\label{eq13}
\end{equation}
The following analysis is in the spirit of the quasiclassical method of Wentzel, Kramers and Brillouin (WKB), see \cite{Si13} for more details. A key feature of the WKB method is a notion of a turning point of a classical trajectory, which in the context of (10) is given by $r_t=|n-\mu|/k$. The two linearly independent solutions at $r>r_t$, $\psi_{n,{\rm out}}^{(+)}(r)$ and $\psi_{n,{\rm out}}^{(-)}(r)$, are determined by asymptotics 
$$
\psi_{n,{\rm out}}^{(\pm)}(r)\,\,\,
{\sim_{_{_{\!\!\!\!\!\!\!\!\!\! r\rightarrow \infty}}}}{\frac{1}{\sqrt{kr}}}e^{\pm{\rm i}kr},
$$
while the two linearly independent solutions at $r<r_t$, $\psi_{n,{\rm in }}^{(+)}(r)$ and $\psi_{n,{\rm in}}^{(-)}(r)$, are determined by asymptotics
$$
\psi_{n,{\rm in}}^{(\pm)}(r)\,\,\,{\sim_{_{_{\!\!\!\!\!\!\!\!\! r \rightarrow 0}}}} (kr)^{\mp|n-\mu |}.
$$
We are interested in the asymptotics of the solution to (3) at large distances from the vortex, hence $r>r_t$. However, matching point $r_c$ can be either larger (for modes with $|n-\mu|<kr_c$), or smaller (for modes with $|n-\mu|>kr_c$) than the turning point. In view of this, we obtain at $r \gg k^{-1}$: 
\begin{equation}
\psi(r,\varphi)=\psi^{(0)}(r,\varphi)+\psi^{(c)}(r,\varphi),\label{eq14}
\end{equation}
where
\begin{equation}
\psi^{(0)}(r,\varphi)=\sum\limits_{n\in \mathbb{Z}}e^{{\rm i}n\varphi}a_n[\psi_{n,{\rm out}}^{(+)}(r)+\psi_{n,{\rm out}}^{(-)}(r)]\label{eq15}
\end{equation}
and
\begin{eqnarray}
\psi^{(c)}(r,\varphi)=-\sum\limits_{|n-\mu|\leq kr_c}e^{{\rm i}n\varphi}a_n[1+C_n(r_c)]\psi_{n,{\rm out}}^{(+)}(r)-\nonumber \\
-\sum\limits_{|n-\mu|>kr_c}e^{{\rm i}n\varphi}a_nC_n(r_c)\psi_{n,{\rm out}}^{(+)}(r),\label{eq16}
\end{eqnarray}
\begin{equation}
C_n(r_c)=\left.\frac{W(\psi_{n,{\rm out}}^{(-)},\,\tau_n)+\kappa\psi_{n,{\rm out}}^{(-)}\tau_n}{W(\psi_{n,{\rm out}}^{(+)},\,\tau_n)+\kappa\psi_{n,{\rm out}}^{(+)}\tau_n}\right|_{r=r_c},\qquad |n-\mu|\leq kr_c,\label{eq17}
\end{equation}
\begin{equation}
C_n(r_c)=\left.\frac{W(\psi_{n,{\rm in}}^{(-)},\,\tau_n)+\kappa\psi_{n,{\rm in}}^{(-)}\tau_n}{W(\psi_{n,{\rm in}}^{(+)},\,\tau_n)+\kappa\psi_{n,{\rm in}}^{(+)}\tau_n}\right|_{r=r_c},\qquad |n-\mu|> kr_c.\label{eq18}
\end{equation}

Choosing the direction of the incoming wave as $\varphi=\pm\pi$, we impose condition 
\begin{equation}
\lim\limits_{r\rightarrow\infty}e^{{\rm i}kr}\psi(r,\,\pm\pi)=1,\label{eq19}
\end{equation}
which determines coefficient $a_n$ as 
\begin{equation}
a_n=\frac{1}{\sqrt{2\pi}}e^{{\rm i}(|n|-\frac 12|n-\mu|)\pi}.\label{eq20}
\end{equation}
In view of this and asymptotics 
\begin{equation}
\psi_{n,{\rm out}}^{(\pm)}(r)=\frac{1}{\sqrt{kr}}\exp[\pm{\rm i}(kr-\frac{1}{2}|n-\mu|\pi-\frac 14\pi)],\qquad r\rightarrow \infty,\label{eq21}
\end{equation}
we obtain
\begin{equation}
\psi^{(0)}(r,\varphi)=e^{{\rm i}kr \cos\varphi}e^{{\rm i}\mu[\varphi-{\rm sgn}(\varphi)\pi]}+\frac{{\rm i}\sin(\mu\pi)}{\sqrt{2\pi k}}\frac{e^{{\rm i}([\![\mu]\!]+\frac 12)\varphi}}{\sin(\varphi/2)}
\frac{e^{{\rm i}(kr+\pi/4)}}{\sqrt{r}},\label{eq22}
\end{equation}
where it is implied that $-\pi<\varphi<\pi$, the sign function is ${\rm sgn}(u)=\pm 1$ at $u\gtrless 0$, and $[\![u]\!]$ denotes the integer part of quantity $u$ (i.e. the integer which is less than or equal to $u$). On the right-hand side of (22), the first term is an incoming wave, while the second term is an outgoing cylindrical wave, $r^{-1/2}\exp[{\rm i}(kr+\pi/4)]$, times a $\varphi$- and $k$-depending factor. This factor is the famous Aharonov-Bohm scattering amplitude \cite{Aha} which corresponds to the case when the internal structure of the magnetic vortex is neglected and the limit of $r_c\rightarrow 0$ is taken. The squared absolute value of the amplitude yields the Aharonov-Bohm scattering differential cross section: 
\begin{equation}
\frac{\rm d\sigma^{(AB)}}{{\rm d} z\rm d\varphi}=\frac{1}{2\pi k}\frac{\sin^2(\mu \pi)}
{\sin^2(\varphi/2)}.\label{eq23}
\end{equation}
 
Meantime, the $r_c$-dependent part of the wave function, $\psi^{(c)}(r,\varphi)$ can be presented in the following form:
\begin{equation}
\psi^{(c)}(r,\varphi)=[f_1(k,\varphi)+f_2(k,\varphi)+f_3(k,\varphi)]
\frac{e^{{\rm i}(kr+\pi/4)}}{\sqrt{r}},\label{eq24}
\end{equation}
where
\begin{equation}
f_1(k,\varphi)=\frac{\rm i}{\sqrt{2\pi k}}\sum\limits_{|n-\mu|\leq kr_c}
e^{{\rm i}n\varphi}e^{{\rm i}(|n|-|n-\mu|)\pi},\label{eq25}
\end{equation}
\begin{equation}
f_2(k,\varphi)=\frac{\rm i}{\sqrt{2\pi k}}\sum\limits_{|n-\mu|\leq kr_c}
e^{{\rm i}n\varphi}e^{{\rm i}(|n|-|n-\mu|)\pi}\,C_n(r_c),\label{eq26}
\end{equation}
\begin{equation}
f_3(k,\varphi)=\frac{\rm i}{\sqrt{2\pi k}}\sum\limits_{|n-\mu|> kr_c}
e^{{\rm i}n\varphi}e^{{\rm i}(|n|-|n-\mu|)\pi}\,C_n(r_c).\label{eq27}
\end{equation}
If the absolute value of $\kappa$ exceeds considerably the absolute value of the Wronskian (divided by the product of its arguments) in (17) and (18), then this corresponds to the formal limit of $|\kappa|\rightarrow\infty$. In this limit the case of an impenetrable magnetic vortex with the Dirichlet boundary condition at its edge is recovered, and this case has been comprehensively analysed elsewhere, see \cite {Si10,Si11}. Otherwise, i.e. if the Wronskian in (17) and (18) is essential, we deal with the case of a penetrable magnetic vortex. It should be noted in the first place that amplitude $f_1$ (25) is the same both for the cases of penetrability and impenetrability, being independent both of the magnetic field distribution inside the vortex and of the choice of a boundary condition at the vortex edge; this is the amplitude of the Fraunhofer diffraction which is strongly peaked in the forward ($\varphi=0$) direction \cite{Si10,Si11}. Amplitude $f_3$ (27) is estimated to be of order of $\sqrt{r_c}O[(kr_c)^{-1/6}]$ (details will be published elsewhere). Thus, both $f_3$ and the Aharonov-Bohm scattering amplitude (which is of order of $\sqrt{r_c}O[(kr_c)^{-1/2}]$, see (22)) are negligible as compared to $f_1$ and $f_2$ which are of order of $\sqrt{r_c}$. The most cumbersome task is the calculation of the sum in (26). The asymptotics in the form of (21) does not suffice, and we use the following asymptotics which is obtainable in the WKB approximation:
\begin{equation}
\psi^{(\pm)}_{n,{\rm out}}(r)=[kr\dot{\xi}_n(kr)]^{-1/2}\exp\left\{{\pm {\rm i}}[\xi_n(kr)-\pi/4]\right\},\label{eq28}
\end{equation}
where
\begin{equation}
\xi_n(y)=\int\limits_{|n-\mu|}^{y}{\rm d}u\sqrt{1-\left(\frac{n-\mu}{u}\right)^2},\label{eq29}
\end{equation}
and $\dot{\xi}_n(y)\equiv\partial_y\xi_n(y)$ is the integrand in (29). Choosing $\tau_n(r)$ to be real, we get its asymptotics in the WKB approximation: 
\begin{equation}
\tau_n(r)=[kr\dot{\zeta}_n(kr)]^{-1/2}\cos[\zeta_n(kr)-\pi/4],\label{eq30}
\end{equation}
where
\begin{equation}
\zeta_n(y)=\int\limits_{y_0}^{y}{\rm d}u\sqrt{1+\frac{2\mu\sigma}{(kr_c)^2}-\left[\frac{n-\gamma(u)}{u}\right]^2},\label{eq31}
\end{equation}
$\dot{\zeta}_n(y)\equiv \partial_y \zeta_n(y)$, $\sigma=1$ ($\sigma=-1$) for the upper (lower) component of $\tau_n(r)$, $y_0$ is a zero of the integrand in (31), which is the closest from the left to point $y=kr_c$. We take account for the fact that $(kr_c)^2\gg 2|\mu|$, see (2), then the dependence on $\sigma$ in (31) (and, consequently, on spin in $\tau_n$ (30)) drops out. In view of $\gamma(kr_c)=\mu$, we get 
\begin{equation}
\dot{\zeta}_n(kr_c)=\dot{\xi}_n(kr_c)=\sqrt{1-\left(\frac{n-\mu}{kr_c}\right)^2}.\label{eq32}
\end{equation}
A careful analysis reveals that amplitude $f_2$ (26) vanishes in the forward direction as $O(\sqrt{|\varphi|})$. Thus, the intefrerence between the amplitudes, $f_1f_2^*+f_1^*f_2$, is absent, and the differential cross section of the scattering process consists of two terms: 
\begin{equation}
\frac{{\rm d}\sigma}{{\rm d}z{\rm d}\varphi}=\frac{{\rm d}\sigma_1}{{\rm d}z{\rm d}\varphi}+\frac{{\rm d}\sigma_2}{{\rm d}z{\rm d}\varphi},\label{eq33}
\end{equation}
where ${\rm d}\sigma_{1,2}/({\rm d}z{\rm d}\varphi)=|f_{1,2}|^2$. Note that the differential cross section of the Fraunhofer diffraction is \cite{Si10,Si11}
\begin{equation}
\frac{{\rm d}\sigma_1}{{\rm d}z{\rm d}\varphi}=\frac{2}{\pi k}\frac{\sin^2(kr_c\varphi/2)}{\sin^2(\varphi/2)}
\cos^2(\mu\pi+kr_c\varphi/2)\qquad (-\pi<\varphi<\pi).\label{eq34}
\end{equation}

Following further in the analysis of amplitude $ f_2$ (26), we note that, owing to its vanishing in the forward direction, the contribution of modes with $|n-\mu|\approx kr_c$ is inessential. This means that, while substituting the WKB asymptotics (28) and (30) into $C_n$ (17), one has to take account for the fact that $[\dot{\zeta}_n(kr_c)]^{-1}$ is bounded from above, and, therefore, 
\begin{equation}
\left|\tan\left[\zeta_n(kr_c)-\pi/4\right]\right|\gg \left|\frac{\kappa}{kr_c\dot{\zeta}_n(kr_c)}-\frac{(n-\mu)\dot{\gamma}(kr_c)}{2(kr_c)^2\dot{\zeta}_n^3(kr_c)}\right|.\label{eq35}
\end{equation}
As a result, we obtain 
\begin{equation}
f_2(k,\varphi)=-\frac{{\rm i}}{\sqrt{2\pi k}}\sum\limits_{|n-\mu|\leq kr_c}e^{{\rm i}[n\varphi+\mu{\rm sgn}(n-\mu)\pi]}e^{2{\rm i}[\zeta_n(kr_c)-\xi_n(kr_c)]}.\label{eq36}
\end{equation}

To calculate the sum in (36) in asymptotics (1), we use the Poisson summation formula, 
\begin{equation}
\sum\limits_{|n-\mu|\leq s}e^{{\rm i}\chi(n,s)}=\sum\limits_{l\in \mathbb{Z}}\int\limits_{-s_-}^{s_+}{\rm d}n\exp\left\{{\rm i}[\chi(n,s)-2\pi nl]\right\}+  \frac 12e^{{\rm i}\chi(s_+,s)}+
\frac 12e^{{\rm i}\chi(-s_-,s)},\label{eq37}
\end{equation}
where $s_\pm=[\![s\pm\mu]\!]$. If $s_++s_-\gg 1$ and $\chi(n,s)$ is convex upwards (downwards), $\frac{\partial^2}{\partial n^2}\chi(n,s)<0$ $\left(\frac{\partial^2}{\partial n^2}\chi(n,s)>0\right)$, on interval $-s_-<n<s_+$, then only a finite number of terms in the series on the right-hand side of (37) contribute to the leading asymptotics at $s\gg 1$, and one can use the method of stationary phase for its evaluation. Namely, if equation 
\begin{equation}
\frac{\partial}{\partial n}\chi(n,s)|_{n=n_j}-2\pi l_j=0 \label{eq38}
\end{equation}
determines a stationary point inside the interval, $-s_-<n_j<s_+$, for some values of $l$ denoted by $l_j$, then 
\begin{equation}
\sum\limits_{|n-\mu|\leq s}e^{{\rm i}\chi(n,s)}=\sum\limits_{l_j}\exp\left\{{\rm i}[\chi(n_j,s)-2\pi n_jl_j]\right\} \sqrt{\frac{2\pi e^{\mp{\rm i}\pi/2}}{\mp \left.\left[\frac{\partial^2}{\partial n^2}\chi(n,s)\right]\right|_{n=n_j}}} + O(1),\label{eq39}
\end{equation}
where the upper and lower signs correspond to the cases of convexity upwards and downwards, respectively.

In principle, for some $l_j$, there may be several well separated stationary points, $n_{jJ}$ ($J=1,\ldots,N$), each one corresponding to an interval of definite convexity, with neighbouring intervals of opposite convexities; a generalization to such a case is obvious. However, as the neghbouring stationary points, say, $n_{jJ}$ and $n_{jJ+1}$ come close together, their contributions to (39) diverge, since there is a point between them, $n_{jJ0}$, $n_{jJ}<n_{jJ0}<n_{jJ+1}$, where
\begin{equation}
\left.\left[\frac{\partial^2}{\partial n^2}\chi(n,s)\right]\right|_{n=n_{jJ0}}=0.\label{eq40}
\end{equation}
The expansion of phase $\chi(n,s)$ in the vicinity of $n_{jJ0}$ is 
\begin{equation}
\chi(n,s)=\chi(n_{jJ0},s)+\alpha_{jJ1}(n-n_{jJ0})+\frac{1}{3!}\alpha_{jJ3}(n-n_{jJ0})^3,\label{eq41}
\end{equation}
where
\begin{equation}
\alpha_{jJ1}=\left.\left[\frac{\partial}{\partial n}\chi(n,s)\right]\right|_{n=n_{jJ0}},\qquad \alpha_{jJ3}=\left.\left[\frac{\partial^3}{\partial n^3}\chi(n,s)\right]\right|_{n=n_{jJ0}}.\label{eq42}
\end{equation}
Thus, in the case when the neighbouring stationary points are not well separated, there is an additional contribution to the sum in (39), which can be written in the following form (see, e.g., \cite{New}):
$$
\exp\left\{{\rm i}\left[\chi(n_{jJ0},s)-2\pi n_{jJ0}l_j\right]\right\}2\pi\left(\frac{2}{|\alpha_{jJ3}|}\right)^{1/3} {\rm Ai}\left[{\rm sgn}(\alpha_{jJ3})\alpha_{jJ1}\left(\frac{2}{|\alpha_{jJ3}|}\right)^{1/3}\right],
$$
where ${\rm Ai}(y)=\pi^{-1}\int\limits_{0}^{\infty}{\rm d}u\,\cos(yu+\frac 13 u^3)$ is the Airy function (see, e.g., \cite{Abra}) which is exponentially damped at real positive values of its argument, while oscillating at real negative ones. 

Concerning the sum in $f_2$ (36), we have
\begin{equation}
\chi(n,kr_c)=n\varphi+\mu{\rm sgn}(n-\mu)\pi+2[\zeta_n(kr_c)-\xi_n(kr_c)], \label{eq43}
\end{equation}
and the condition that the phase be stationary takes form
\begin{equation}
\frac{\partial}{\partial n}[\xi_n(kr_c)-\zeta_n(kr_c)]|_{n=n_j}=\frac{1}{2}\varphi-l_j\pi. \label{eq44}
\end{equation}
In the case of the uniform magnetic field one has 
\begin{equation}
\gamma(y)=\mu y^2/(kr_c)^2\label{eq45}
\end{equation}
and
\begin{eqnarray}
\zeta_n(kr_c)=\frac 12\sqrt{(kr_c)^2-(n-\mu)^2}+\nonumber \\
+\frac{1}{4|\mu|}[(kr_c)^2+2\mu n]\arccos\frac{(kr_c)^2+2\mu(n-\mu)}{kr_c\sqrt{(kr_c)^2+4\mu n}}-\nonumber \\ -\frac{|n|}{2}\arccos\frac{-(kr_c)^2+2(n-\mu)n}{kr_c\sqrt{(kr_c)^2+4\mu n}}.\label{eq46}
\end{eqnarray}
Note also obvious relation, see (29),
\begin{equation}
\xi_n(kr_c)=\sqrt{(kr_c)^2-(n-\mu)^2}-|n-\mu|\arccos\frac{|n-\mu|}{kr_c}.\label{eq47}
\end{equation}
Then amplitude $f_2$ (36) in asymptotics (1) can be calculated with the use of (39)-(44) (details will be published elsewhere). We present here the results for the differential cross section:
\begin{equation}
\frac{{\rm d}\sigma_2}{{\rm d}z{\rm d}\varphi}=r_c|\sin\frac{\varphi}{2}|\Biggl[ \frac{1}{2}\frac{1+\left(\frac{kr_c}{2\mu}\right)^2\cos \varphi}{\sqrt{1-\left(\frac{kr_c}{2\mu}\right)^2\sin^2\frac{\varphi}{2}}}-{\rm sgn}(\varphi)\frac{kr_c}{2\mu}\cos\frac{\varphi}{2}\Biggr],\, 2|\mu| > kr_c,\label{eq48}
\end{equation}
\begin{eqnarray}
\frac{{\rm d}\sigma_2}{{\rm d}z{\rm d}\varphi}=\frac{r_c|\sin\frac{\varphi} {2}|}{\sqrt{1-\left(\frac{kr_c}{2\mu}\right)^2\sin^2\frac{\varphi}{2}}}\Biggl\{1+\left(\frac{kr_c}{2\mu}\right)^2\cos\varphi +\left[\left(\frac{kr_c}{2\mu}\right)^2-1\right] \times \nonumber\\ 
\times \sin\left[4|\mu|\arccos\left(\frac{kr_c}{2|\mu|}|\sin\frac{\varphi}{2}|\right)
-2kr_c|\sin\frac{\varphi}{2}|\sqrt{1-\left(\frac{kr_c}{2\mu}\right)^2\sin^2\frac{\varphi}{2}}\right]\Biggr\}, \nonumber \\ 
0 \leq -{\rm sgn}(\mu)\varphi < 2\arcsin\frac{2|\mu|}{kr_c}, \quad 2|\mu| \leq kr_c,\label{eq49}
\end{eqnarray}
\begin{eqnarray}
\frac{{\rm d}\sigma_2}{{\rm d}z\,{\rm d}\varphi}=(2\pi)^2 (2|\mu|)^{4/3}\left[\left(\frac{kr_c}{2\mu}\right)^{2}-1\right] \times \nonumber \\ 
\times {\rm Ai}^2\left[-{\rm sgn}(\mu)\left(\varphi+2{\rm arcsin}\frac{2\mu}{kr_c}\right)(2|\mu|)^{2/3} \sqrt{\left(\frac{kr_c}{2\mu}\right)^{2}-1}\right], \nonumber \\ 
\varphi\approx -2{\rm arcsin}\frac{2\mu}{kr_c},\qquad 2|\mu| \leq kr_c.\label{eq50}
\end{eqnarray}

\section{Discussion of results and conclusion}

In the present paper, we have considered scattering of a nonrelativistic charged particle with spin by a penetrable magnetic vortex of a transverse size which exceeds considerably the particle wavelength, see (1). The values of the vortex flux are subject to condition (2) which includes all possibilities for attaining steady magnetic fields in the present-day (and even future) laboratory facilities. Let us first stress on some general features which are independent of the details of the magnetic field distribution inside the vortex.

The differential cross section of the scattering process consists of two terms, see (33): a first one describes the Fraunhofer diffraction of the particle on the edge of the vortex, see (34), and a second one describes a penetration of the particle through the vortex. The absence of interference between the effects of diffraction and penetration is due to the fact that the diffraction effect contributes exclusively in the forward direction where the penetration effect gives no contribution. The diffraction effect is the same for both cases of the vortex penetrability and impenetrability \cite{Si12}. As long as the vortex is penetrable ($|\kappa|<\infty$), the penetration effect is independent of a measure of penetrability, i.e. of the value of $\kappa$. Both the diffraction and penetration effects are independent of the particle spin.

A further analysis has been performed for the case of a uniform distribution of the magnetic field inside the vortex. For a strong magnetic field, $|B|>\hbar kc/(|e|r_c)$, the differential cross section of the penetration effect is given by (48) which coincides with the scattering differential cross section obtainable in the framework of classical theory,
\begin{equation}
\frac{{\rm d}\sigma^{(\rm class)}}{{\rm d}z\,{\rm d}\varphi}=\left|\sin\frac \varphi 2\right|\Biggl[\frac 12\frac{r_c^2+r_B^2\cos\varphi}{\sqrt{r_c^2-r_B^2\sin^2\frac{\varphi}{2}}}-{\rm sgn}(eB\varphi)r_B\cos\frac{\varphi}{2}\Biggr],\quad r_B<r_c,\label{eq51}
\end{equation}
where we recall that $r_B=\hbar kc/|eB|$ is the radius of the classical particle orbit in the uniform magnetic field. The particle is deflected at all angles, $-\pi<\varphi<\pi$, but the cross section is asymmetric with respect to the forward ($\varphi=0$) direction, as well as with respect to the change of the sign of the magnetic field or the particle charge; the cross section is symmetric under the simultaneous change, $\varphi\rightarrow-\varphi$ and $eB\rightarrow-eB$.

For a weak magnetic field, $|B| \leq \hbar kc/(|e|r_c)$, the differential cross section of the penetration effect is given by (49) and (50). Let us compare the latter with the appropriate result in classical theory,
\begin{equation}
\frac{{\rm d}\sigma^{(\rm class)}}{{\rm d}z\,{\rm d}\varphi}=|\sin\frac{\varphi}{2}|\frac{r_c^2+r_B^2\cos\varphi}{\sqrt{r_c^2-r_B^2\sin^2\frac\varphi 2}},\quad 0 \leq -{\rm sgn}(eB)\varphi \leq 2{\rm arcsin}\frac{r_c}{r_B},\quad r_B \geq r_c.\label{eq52}
\end{equation}
According to (49) and (50), the particle is deflected only to one side from the forward direction, within the angle range restricted to $-2 {\rm arcsin}[eBr_c/(\hbar kc)] \lessapprox \varphi\leq 0$ in the case of $eB>0$ or to $0\leq\varphi \lessapprox -2{\rm arcsin}[eBr_c/(\hbar kc)]$ in the case of $eB<0$. This is a manifestation of the fact that, in classical treatment, the scattering angle is not a monotonic function of the impact parameter, but has an extremum, $\varphi=\varphi_{\rm extr}$, where $\varphi_{\rm extr} = -2{\rm arcsin}[eBr_c/(\hbar kc)]$. Consequently, there are two branches of the values of the impact parameter, giving the same values of the scattering angle within the restricted range; the values of the scattering angle out of the range are not accessible at any of the values of the impact parameter. Quantum-mechanically, there is an interference between these two branches, resulting in the last oscillating term in figure brackets in (49). The interference disappears at $|B| = \hbar kc/(|e|r_c)$, when the classical and quantum-mechanical treatments give the same result 
\begin{equation}
\frac{{\rm d}\sigma^{(\rm class)}}{{\rm d}z\,{\rm d}\varphi}=r_c |\sin \varphi|,\quad 0 \leq -{\rm sgn}(eB)\varphi \leq \pi,\quad r_B = r_c.\label{eq53}
\end{equation} 
Otherwise, there is one more distinction in addition to the already mentioned interference.
Classically, the cross section diverges at $\varphi=\varphi_{\rm extr}$, see (52), but, quantum-mechanically, it is regulated by the Airy function, see (50). In general, such an effect is well-known in literature (see, e.g., \cite{New}), giving rise to the rainbow phenomenon in meteorology.

Since the AB effect is an essentially quantum effect that is alien to classical physics, it may manifest itself only in situations when there are distinctions between classical and quantum treatments. Rainbow, see (50), leaves no room for the AB effect, but the effect might be visible in interfence between two branches of the classical impact parameter: there are oscillations which are periodic in the value of flux $\Phi$ with period $2\pi\hbar c/|e|$ (London flux quantum) in the differential cross section of the penetration effect in the case of a weak magnetic field, see (49). The AB effect (periodicity in $\Phi$) disappears at $\varphi=\varphi_{\rm extr}$, whereas it is maximally exposed at $\varphi=0$. However, namely in the vicinity of the forward direction, the cross section is damped due to factor $|\sin \frac\varphi 2|$, see (49), and this makes the detection of the AB effect via such a way to be somewhat problematic.

Much more favourable prospects are for the detection of the AB effect via the Fraunhofer diffraction which is a purely quantum phenomenon. This issue was discussed in detail in \cite{Si12,Si13}, and we only note here that the diffraction pattern in the differential cross section comprises one or two peaks with their shapes and positions changing periodically in the value of flux $\Phi$ with period $2\pi \hbar c/|e|$, see (34). 

A scattering event corresponds to an infinitely large distance between the target (magnetic vortex) and the detector. As the distance decreases, the Fraunhofer diffraction gives way to the Fresnel diffraction, i.e. the diffraction in converging beams, see, e.g., \cite{Bo}. The number of the diffraction peaks on the detection screen increases, and this is a pattern which is usually observed in the interference experiments aiming at the verification of the AB effect. We conclude that the physical cause of the emergence of the interference pattern in such experiments is diffraction of electron beams passing by the region of the magnetic flux from different sides. As follows from scattering theory, the electrons penetrating into the region of the magnetic flux are deflected away from the detection screen which is placed in the forward direction. That is why the penetrability or impenetrability of the magnetic vortex has no impact on the flux-dependent fringe shift in the interference pattern. In view of this, the first simple experiment with iron whiskers at the H.H. Wills Physics Laboratory of the University of Bristol \cite{Cha} confirms the AB effect at a no less extent than the later much more elaborate experiments (see \cite{Pes,Ton}) involving toroidal magnets enclosed in shells of conducting and superconducting materials.

\section*{Acknowledgments}

The work was supported by the National Academy of 
Sciences of Ukraine (project No.0112U000054), by the the Program 
of Fundamental Research of the Department of Physics and Astronomy 
of the National Academy of Sciences of Ukraine (project No.0112U000056) 
and by the ICTP -- SEENET-MTP grant PRJ-09 ``Strings and Cosmology''.

\end{document}